\documentstyle[A4,11pt,epsfig,twoside]{article}
\begin{document}
\thispagestyle{empty}
\begin{flushright}
                                                   CERN-TH/2002-337\\
                                                   IISc/CTS/12-02\\
                                                   hep-ph/0212191  \\
\end{flushright}
\vspace{0.5cm}
\begin{center}
{\Large
{\bf Probing the CP  of the Higgs at a $\gamma \gamma$ collider using
        $\gamma \gamma \rightarrow t \bar t \rightarrow lX$.
       \footnote{Talk presented by A. de Roeck at the International Linear
        Collider Workshop, Jeju Island, Aug. 26-30,2002.} }}\\[5ex]

R.M. Godbole$^{a,}$  
       \footnote{Permanent Address: CTS, Indian Institute of Science,
                 Bangalore, 560 012, India}   
 S. D. Rindani $^{b}$ 
   and  
  R. K. Singh $^{c}$  
\\[5ex]
$^a$ {\it  CERN, Theory Division, CH-1211, Geneva 23, Switzerland}
\\
$^b$ {\it Physical Research Laboratory, Ahmedabad, 380 009
            India} \\      
$c$ {\it CTS, Indian Institute of Science, Bangalore, 560 012, India}

\end{center}
%
{\begin{center}
ABSTRACT
\vspace{0.5cm}

\parbox{13cm}{
We present results of an investigation to study CP violation in the
Higgs sector in  $t\bar t$ production at a $\gamma\gamma$-collider.
This is done in a model independent way in terms of six form-factors
$\{\Re(S_{\gamma}), \Im(S_{\gamma}), \Re(P_{\gamma}),
\Im(P_{\gamma}), S_t, P_t\}$  which parameterize the CP mixing in Higgs
sector. The angular distribution of the decay lepton from $t/\bar t$ is
shown to be independent of any CP violation in the $tbW$ vertex.
Hence it can be used as a diagnostic of the CP mixing. We study how well
one can probe different combinations of the form factors by measurements
of the combined asymmetries that we construct, in the initial state
lepton (photon) polarization and the final state lepton charge, using
only circularly polarized photons. We show that the method
can be sensitive to loop-induced CP violation in the Higgs sector in the
MSSM.
}
\end{center}}
\newpage     
\title{Probing the CP  of the Higgs at a $\gamma \gamma$ collider using 
        $\gamma \gamma \rightarrow t \bar t \rightarrow lX$. 
       \thanks{Talk presented by A. de Roeck at the International Linear 
        Collider Workshop, Jeju Island, Aug. 26-30, 2002.}} 

\author{R.M. Godbole$^1$
        \thanks{e-mail address: rohini.godbole@cern.ch}~ \thanks{Permanent
          Address: CTS, Indian Institute of Science,
                 Bangalore, 560 012, India}~~
        S. D. Rindani $^2$
            \thanks{e-mail address: saurabh@prl.ernet.in}~~ and ~~
        R.  K. Singh $^3$ \thanks{e-mail address: ritesh@cts.iisc.ernet.in}
\\
        $^1$ {\it  CERN, Theory Division, CH-1211, Geneva 23, Switzerland}
\\
        $^2$ {\it Physical Research Laboratory, Ahmedabad, 380 009
            India}
\\      $3$ {\it CTS, Indian Institute of Science,
                 Bangalore, 560 012, India}}

\date{}
\maketitle
\begin{abstract}
We present results of an investigation to study CP violation in the 
Higgs sector in  $t\bar t$ production at a $\gamma\gamma$-collider.  
This is done in a model independent way in terms of six form-factors
$\{\Re(S_{\gamma}), \Im(S_{\gamma}), \Re(P_{\gamma}),
\Im(P_{\gamma}), S_t, P_t\}$  which parameterize the CP mixing in Higgs 
sector. The angular distribution of the decay lepton from $t/\bar t$ is 
shown to be independent of any CP violation in the $tbW$ vertex. 
Hence it can be used as a diagnostic of the CP mixing. We study how well 
one can probe different combinations of the form factors by measurements 
of the combined asymmetries that we construct, in the initial state 
lepton (photon) polarization and the final state lepton charge, using 
only circularly polarized photons. We show that the method 
can be sensitive to loop-induced CP violation in the Higgs sector in the 
MSSM.
\end{abstract}
\section{Introduction}
While the standard model (SM) has been proved to provide the correct 
description of fundamental particles and their interactions, direct 
experimental verification of the Higgs sector and a basic understanding 
of the mechanism for the generation of the observed CP violation is 
still lacking.
Many models with an extended Higgs sector have  CP violation in the Higgs 
sector.
In this context  there are then two important questions that need to be 
answered viz., if CP is conserved in the Higgs sector, how well can the 
CP transformation properties of the, possibly more than one, neutral Higgses 
be established and  if it is violated how is this reflected in Higgs mixing 
and the couplings.  The CP violation in the Higgs sector can be 
either explicit, spontaneous or loop-induced. The last has been studied in the
context of the minimal Supersymmetric Standard Model (MSSM)  in great detail
recently and arises from loops containing sparticles and nonzero phases of
the MSSM parameters $\mu$ and $A_t$.

$\gamma \gamma$ colliders are shown to make possible  an accurate 
determination of the $\gamma\gamma$ width of the Higgs and
allow possibilities  of search for the $H/A$ of  the MSSM  at points 
in parameter space not accessible to LHC. More importantly they 
provide unique opportunity of determination of CP properties of Higgs using  
polarized  photon beams by studying the dependence of the cross-section
on the initial beam polarization~\cite{gunion} as well as the polarization
of the $t/ \bar t$ produced in the final state~\cite{asakawa,asak1} in the Higgs
decay. The last has been studied in a model independent way. In this talk
we present results of our studies~\cite{us} of the  
$\gamma \gamma \rightarrow \phi \rightarrow t \bar t$  production
followed by decay of the polarized $t$ and develop a strategy to
determine the CP properties of  the $\phi$ couplings,  by probing the $t$
polarization through the decay $l$ distributions, for which analytical
expressions were obtained. Since the top decays rapidly enough 
the angular distributions of the lepton coming from the top decay
can provide a good probe of the initial top polarization and has been 
shown to work effectively in the analysis of top dipole moment~\cite{prds1} 
and  CP-violating $\gamma\gamma Z$ coupling~\cite{plbs2}.

\section{Formalism and Decay $l$ angular distribution}
We perform our calculations in a model independent way. We
parametrise the vertices ${\cal V}_{t\bar t \phi},{\cal V}_{\gamma\gamma\phi}$
of the scalar $\phi$ with a $t \bar t$ and $\gamma \gamma $ pair, 
in a manner similar to Ref.~\cite{asakawa} as
\begin{equation}
{\cal V}_{t\bar t \phi}  = -ie\frac{m_t}{M_W} \left(S_t+i\gamma^5P_t\right)
\label{eq:vert1}
\end{equation}
and 
\begin{equation}
{\cal V}_{\gamma\gamma\phi}=  \frac{-i\sqrt{s}\alpha}{4\pi}\left[S_{\gamma}(s)
\left(\epsilon_1.\epsilon_2-\frac{2}{s}(\epsilon_1.k_2)(\epsilon_2.k_1) \right)
- P_{\gamma}(s)\frac{2}{s}\epsilon_{\mu\nu\alpha\beta}\epsilon_1^{\mu}
\epsilon_2^{\nu} k_1^{\alpha} k_2^{\beta} \right].
\label{eq:vert2}
\end{equation}
$k_1$ and $k_2$ are the four-momenta of colliding photons and 
$\epsilon_{1,2}$ are photon polarization vectors.  $S_t, P_t$ can be 
real constants without loss of generality whereas  $S_\gamma, P_\gamma$ 
are complex form factors. Simultaneous presence of $P_t$ and $S_t$ and/or
$S_\gamma$ and $ P_\gamma$ implies CP violation. For the $tbW$ vertex also we 
choose the completely general form
\begin{equation}
\Gamma^{\mu}_{tbW} =-\frac{g}{\sqrt{2}}V_{tb}\left[\gamma^{\mu}
\left(f_{1L}P_L + f_{1R} P_R\right)\right.
-\left.\frac{i}{M_W}\sigma^{\mu\nu}(p_t-p_b)_{\nu}
\left(f_{2L}P_L + f_{2R} P_R\right)\right]
\label{tbwver}
\end{equation}
and similarly for the vertex for  $\bar t$. In the limit of vanishing $b$
masses, and taking $f_{1L}$ to have the SM value 1, the only nonstandard 
part of this vertex which gives non-vanishing contribution, is  $f_{2R}$,
and similarly ${\bar f}_{2L}$ for the vertex of $\bar t$ {\it viz.}  
${\bar\Gamma}^{\mu}_{tbW}$.
For the general $\phi t \bar t ,\phi \gamma \gamma$ and $tbW$ vertex 
given above we use the  helicity amplitudes to calculate the 
analytical expression for differential cross-section for 
$\gamma \gamma \rightarrow t \bar t \rightarrow l^+ b \nu_l \bar t$ and 
hence for the angular distribution of the decay lepton, keeping only the 
linear terms in $f_{2R}, {\bar f}_{2L}$.
 
The differential cross-section is given by a matrix product of the production 
and decay density matrices, integrated over an appropriate phase space.
The production and the decay density matrices $\rho^+(\lambda,\lambda'),
\Gamma(\lambda,\lambda')$ in the $\gamma \gamma $ c.m. frame and in the $t$ 
rest frame respectively are given by,
\begin{eqnarray}
\rho^+(\lambda,\lambda') = e^4\rho'^+(\lambda,\lambda') = {\large \sum}
\rho_1(\lambda_1,\lambda_1')\rho_2(\lambda_2,\lambda_2') 
{\cal M}(\lambda_1,\lambda_2,\lambda,\lambda_{\bar t})
{\cal M}^*(\lambda_1',\lambda_2', \lambda',\lambda_{\bar t})\nonumber \\
\Gamma(\lambda,\lambda') = g^4 |\Delta(p_W^2)|^2  \ \Gamma'(\lambda,
\lambda') = \frac{1}{2\pi}\int d\alpha
 {\large \sum }
M_{\Gamma}(\lambda,\lambda_b,\lambda_{l^+},\lambda_{\nu}) \
M_{\Gamma}^*(\lambda',\lambda_b,\lambda_{l^+},\lambda_{\nu}).
\label{ddecpro}
\end{eqnarray}
Here $M_\Gamma , {\cal M}$ are the decay and production matrix elements,
$\alpha$ : azimuthal angle of $b$-quark in the rest-frame of
$t$-quark with $z$-axis pointing in the direction of momentum of lepton and
$\rho_{1(2)}$ are the photon density matrices.

The decay $l$ angular distribution can be obtained analytically by integrating 
the equation for the differential cross-section 
over $E_l,\cos\theta_t$ and $\phi_l$. It can be shown
that the effect of the anomalous $tbW$ coupling on $l$ angular distribution,
is only an overall factor $1+2r-6\Re(f^\pm)\sqrt{r}$ {\it independent} of 
any kinematical variables. The total width of $t$-quark calculated up-to 
linear order in the anomalous vertex factors receives the {\it same} factor.
Thus to linear approximation in anomalous $tbW$ couplings the angular 
distribution of the decay lepton  unaltered. Hence this is an observable for 
which the only source of the CP-violating asymmetry will be the production 
process\cite{old,us}.

The cross-section $\sigma(\gamma \gamma \rightarrow t \bar t \rightarrow lX)$,
depends on the relative polarizations of the two $\gamma$'s since the $\phi$
exchange diagram contributes only when both colliding photons have same
helicity. Further, the $\gamma \gamma$ collider will be constructed using
laser  backscattered photons and hence the polarization/energy spectrum of
$\gamma$ depends on laser photon and beam lepton ($e^+/e^-$) helicities.
One has to choose $\lambda_e\lambda_l=-1$ to get a hard photon spectrum,
and set $\lambda_{e^-} = \lambda_{e^+}$ to maximize the sensitivity to
possible CP-violating interactions coming from the Higgs coupling. For this 
choice the initial state polarization is completely specified by giving 
the helicity of (say) the $e^-$. In the final state one can look either 
for an $l^+$ or $l^-$. Hence, we have four possible polarized cross-sections:
$\sigma(+,+), \ \sigma(+,-), \ \sigma(-,+), \ \sigma(-,-)$, where the first 
index denotes the helicity of the $e^-$ and second the charge of the lepton.
We wish to construct asymmetries which will be sensitive to a 
CP-violating $\phi$ coupling.

\section{Asymmetries of $\sigma$ w.r.t. initial $\gamma$ polarization and 
final $l$ charge.}
Using these four available cross-section we can now define six asymmetries 
w.r.t. the initial $e^-$ polarization and final $l$ charge as, 
\begin{eqnarray}
{\cal A}_1  =  \frac{\sigma(+,+)-\sigma(-,-)}{\sigma(+,+)+\sigma(-,-)},
{\cal A}_2  =  \frac{\sigma(+,-)-\sigma(-,+)}{\sigma(+,-)+\sigma(-,+)},
{\cal A}_3  =  \frac{\sigma(+,+)-\sigma(-,+)}{\sigma(+,+)+\sigma(-,+)}
\nonumber\\
{\cal A}_4  =  \frac{\sigma(+,-)-\sigma(-,-)}{\sigma(+,-)+\sigma(-,-)},
{\cal A}_5  =  \frac{\sigma(+,+)-\sigma(+,-)}{\sigma(+,+)+\sigma(+,-)},
{\cal A}_6  =  \frac{\sigma(-,+)-\sigma(-,-)}{\sigma(-,+)+\sigma(-,-)}.
\label{asymm}
\end{eqnarray}
Due to the different angular dependence of the  different contributions, 
the $\sigma'$ s are calculated with a cut off on the lepton
angle $\theta_0$, to be optimized to increase sensitivity to CP-violating 
couplings. Out of the six asymmetries, ${\cal A}_1$ and ${\cal A}_2$ 
are purely CP-violating. ${\cal A}_3$ and ${\cal A}_4$ are polarization 
asymmetries for a given lepton charge. ${\cal A}_5$ and ${\cal A}_6$ are 
charge asymmetries  for a given polarization which will  be zero if 
$\theta_0 \rightarrow 0$. Further, only three of these asymmetries are 
linearly independent of each other.

To study these further we choose a specific prediction in 
the MSSM~\cite{asak1} for $\tan \beta =3$, with all sparticles heavy and 
maximal phase. The values we choose are: $m_\phi  = 500 GeV , 
\Gamma_\phi = 1.9 GeV, S_t =  0.33 , P_t =  0.15, S_{\gamma}  =  
-1.3-1.2i , P_{\gamma}  =  -0.51+1.1i.$ We choose beam energy $E_b = 310$ GeV
for this choice of the Higgs mass and the photon spectra, to maximize the 
asymmetries. The asymmetries can be as high as $9 \%$ for (say) ${\cal A}_4$.
Even the  CP-violating asymmetries can be as high as $3$--$4 \%$. The
CP properties of the  Higgs can be determined if one  knows all the {\it six}
form-factors $S_t,  P_t,  \Re(S_{\gamma}),  \Im(S_{\gamma}),  \Re(P_{\gamma}), 
\Im(P_{\gamma})$. These appear in the production density matrix in
eight combinations: the CP-even $x_i$ and CP-odd $y_i$, $(i=1,...4)$
given by $S_t\Re(S_{\gamma}), S_t\Im(S_{\gamma}), P_t\Re(P_{\gamma}),
P_t\Im(P_{\gamma})$ and $S_t\Re(P_{\gamma}), S_t\Im(P_{\gamma}), 
P_t\Re(S_{\gamma}) , P_t\Im(S_{\gamma})$ respectively. The above mentioned 
asymmetries are functions of $x$'s and $y$'s and thus can be used to extract  
information on  these combinations. 

Asymmetries are constructed from the measured cross-sections as,
\begin{equation}
{\cal A}= \frac{\sigma_1-\sigma_2}{\sigma_1+\sigma_2}=
 \frac{\Delta\sigma}{\sigma}.
\end{equation}
The number of events corresponding to the asymmetry are ${\cal L }
\Delta \sigma$. For the asymmetry to be measurable at all we must have at least
${\cal L} \Delta \sigma > f \sqrt{{\cal L} \sigma}$, where $f=1.96 $ for 
95\% c.l. Thus  ${{{\cal L} {\Delta \sigma}}\over {f \sqrt{{\cal L}\sigma}}}  
= {\sqrt{{\cal L}}\over f} \times {{\Delta \sigma} \over {\sqrt{\sigma}}}$
can be taken  as a  measure of the sensitivity. To be more precise, one can 
compare numerical value for a given asymmetry ${\cal A}$ with the 
expected fluctuation in its value at a given level of confidence, viz.,
$\frac{{\cal A}}{\delta{\cal A}} \propto \frac{\Delta\sigma}
{\sqrt{\sigma}}$. 

With this definition of sensitivity we then look for suitable angular cut in 
the lab frame which will  maximize the sensitivity of the measurement.
For ${\cal A}_5, {\cal A}_6$ the optimal choice of the cut off angle s around 
$60^{\circ}$, whereas for the purely CP-violating asymmetries ${\cal A}_1,
{\cal A}_2$ it is $0 ^{\circ}$.  In view of the experimental cut off at small 
angles to the beam direction, we choose two different values of angular cuts;
$20^{\circ}$ and $60^{\circ}$. If for certain values of the form-factors the 
predicted asymmetries lie within the fluctuation from the values expected in 
the SM, then it means that this particular set of values for the form factors
cannot be distinguished from those in the SM at the luminosity we consider. 
We then say that this point falls in the blind region of the 
parameter space. 
In this region the hypothesis that the actual values of the couplings are 
different from the SM expectation cannot be tested.

Thus the set of parameters
$x_i,y_i$ are said to be inside the blind region at a given luminosity if
$$|{\cal A}(\{x_i,y_i\})-{\cal A}_{SM}| \leq \delta{\cal A}_{SM} =
\frac{f}{\sqrt{\sigma_{SM} {\cal L}}}\sqrt{1+{\cal A}_{SM}^2}.$$ 
We took two of the eight possible combinations to be non-zero at a time and
studied how well these can be constrained.

\section{How can the asymmetries be used?}
\begin{figure}[htb]
\rotatebox{270}{ \scalebox{0.55}{\includegraphics{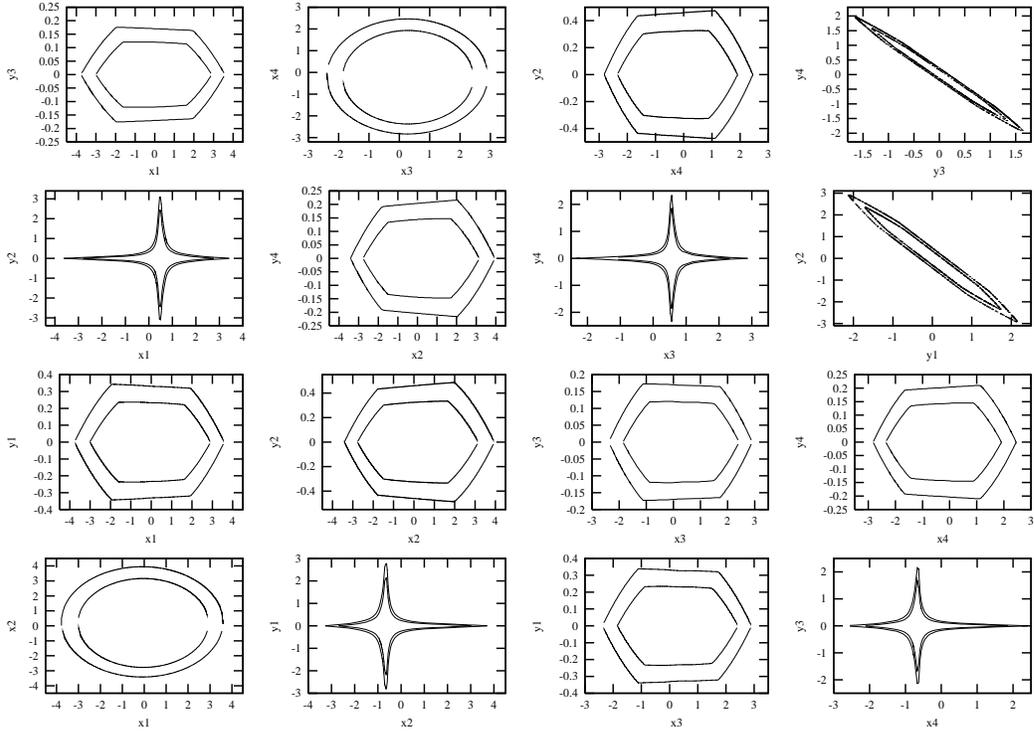}}}
\caption{
The boundaries of blind regions for various pairs of parameters.
Details given in the text}
\end{figure}
Figure 2 shows the boundaries of the blind region as defined above, 
for various $x_i,y_j$ pairs, for luminosity values of  500 and 1000 fb$^{-1}$, 
with beam energy $E_b =$ 310  GeV. Both angular cuts, 
$\theta_0 = 20^{\circ}$ and $60^{\circ}$, are used to put limits at C.L. of 
95\%. The larger region corresponds to 500 fb$^{-1}$, while the smaller 
corresponds to 1000 fb$^{-1}$. We see that  indeed the asymmetries can 
probe for nonzero values of the CP-violating parameters $y_j, j=1,4$. One 
may further ask the question whether it is possible to discriminate a 
particular point in the parameter space of the MSSM predictions against the 
SM as the correct theory. To be able to do that not only is it necessary that 
the particular values of $x_i,y_j$ lie outside the blind region for the 
SM for the pair of parameters under consideration, but further there should be
no overlap of this blind region with that around the values $x_{i}^{mssm}, 
y_{j}^{mssm}$ expected for the MSSM point under consideration. The latter
can be determined again the same way as that for the SM, using   
expected values of the asymmetries for the MSSM point. 
\begin{figure}[htb]
\centerline{
\includegraphics*[scale=0.5]{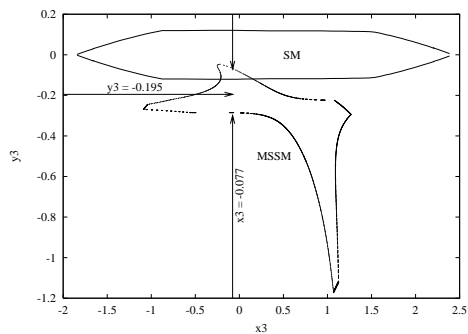}}
\caption{
The boundaries of  blind regions in the parameter space at $95 \%$
c.l. in the $x_3-y_3$ plane, for a luminosity of 1000 fb$^{-1}$ for
$E_b =$ 310 GeV. Both angular cuts, $\theta_0 = 20^{\circ}$ and $60^{\circ}$,
are used for the MSSM point, $x_3=-0.077, \ y_3=-0.195$.\label{fig:smmssm}}
\end{figure}
As Figure 3 shows that one is indeed sensitive to the
values of the parameters predicted in MSSM due to loop effects.
Since in the analysis done above, we hold  values of all the other parameters, 
other than the two being varied, at the values expected in the model
(say the SM) one has to combine  different results of Figure 2
to obtain the range to which the various parameters 
$x_i,y_j, i,j=1,4$ can  be restricted at a given luminosity. The details of
such an analysis are given elsewhere ~\cite{us}. 

Since only three out of the possible six asymmetries are linearly independent 
and there are six independent form factors, it is clear that one needs 
additional information to extract all of them in a completely model independent way. It has been established \cite{asakawa}
that at least {\it in principle} complete determination of the form factors 
using the polarization asymmetries of the final state $t/\bar t$ is possible 
if one uses  linear polarisation of the $\gamma$ along with the circular one.
Our analysis above has studied the possible accuracy of the determination of 
these form factors using the combined asymmetries involving  the
initial lepton (and hence the photon) polarization and the decay lepton
charge, for the case of circular polarization of the initial $\gamma$.
It would be interesting to extend the analysis of the decay $l$ asymmetries,
using linearly polarized $\gamma$.

\section{Summary}
We have studied $\gamma \gamma \rightarrow \phi \rightarrow
t\bar t$ where  $\phi$ is a scalar which may or may not have definite
CP parity. We looked at the process $\gamma \gamma \rightarrow t \bar t
\rightarrow l^\pm X$, where  $l^+/l^-$ comes from decay of $t/\bar t$.
CP-non-conserving  vertices ${\cal V}_{\phi \gamma \gamma}, 
{\cal V}_{t \bar t \phi}$,
can give rise to net polarization asymmetry for the $t$.  The angular
distribution for the decay $l$ is used as an analyzer of $t$ polarization
and hence of CP violation in the Higgs sector. We have studied this in a model
independent way using the $ {\cal V}_{\phi \gamma \gamma}, 
{\cal  V}_{t \bar t \phi}$ 
parametrised in terms of form factors. We first establish that the
decay lepton angular distribution is insensitive
to any anomalous part the $tbW$ coupling $f^{\pm}$ to first order.
We have further constructed  combined asymmetries involving  the
initial lepton (and hence the photon) polarization and the decay lepton
charge. These can put strong limits on CP-violating combinations
of the form factors $y$'s, when only two combinations are varied at a time.
However, the use  of only the circularly polarized photons is found to be
inadequate for simultaneous determination or constraining of all
the form-factors. The analysis thus needs to be extended to include linear
polarization of the photons.

\end{document}